\documentclass[twocolumn,english,showpacs,preprintnumbers,prb]{revtex4}
\usepackage[T1]{fontenc}
\usepackage[latin1]{inputenc}
\usepackage{graphicx}
\usepackage{amssymb}

\makeatletter

\renewcommand{\marginpar}[1]{}

\newcommand{\etal}[1]{, #1}

\usepackage{babel}
\makeatother
\begin{document}
\newcommand{\mubohr}{{\mu_{\mathrm{B}}}}

\newcommand{\efermi}{{\varepsilon_{\mathrm{F}}}}

\newcommand{\bra}[1]{\langle#1|}

\newcommand{\ket}[1]{|#1\rangle}

\newcommand{\ibra}[1]{{}_{\mathrm{I}}\bra{#1}}

\newcommand{\iket}[1]{\ket{#1}_{\mathrm{I}}}

\newcommand{\rotLbl}{{\mathrm{r}}}

\newcommand{\rotbra}[1]{{}_{\rotLbl}\bra{#1}}

\newcommand{\rotket}[1]{\ket{#1}_{\rotLbl}}

\newcommand{\braket}[2]{{\langle#1|#2\rangle}}

\newcommand{\ketbraUnity}[1]{\ket{#1}\bra{#1}}

\newcommand{\braTwo}[2]{\bra{#1,\, \hat{#2}}}

\newcommand{\ketTwo}[2]{\ket{#1,\, \hat{#2}}}

\newcommand{\expect}[1]{\left\langle #1\right\rangle }

\newcommand{\spS}{\ket{S}}

\newcommand{\spup}{\ket{\! \uparrow}}

\newcommand{\spdown}{\ket{\! \downarrow}}

\newcommand{\spupbra}{\bra{\uparrow\! }}

\newcommand{\spdownbra}{\bra{\downarrow\! }}

\newcommand{\spupup}{\ket{\! \uparrow\uparrow}}

\newcommand{\spupdown}{\ket{\! \uparrow\downarrow}}

\newcommand{\spdownup}{\ket{\! \downarrow\uparrow}}

\newcommand{\spdowndown}{\ket{\! \downarrow\downarrow}}

\newcommand{\du}{{\downarrow\uparrow}}

\newcommand{\ud}{{\uparrow\downarrow}}

\newcommand{\rhoElem}[1]{\rho_{#1}}

\newcommand{\rhoDotElem}[1]{{\dot{\rho}_{#1}}}

\newcommand{\rhoDot}{{\rhoDotElem{}}}

\newcommand{\rhoDotuu}{{\rhoDotElem{\uparrow}}}

\newcommand{\rhoDotdd}{{\rhoDotElem{\downarrow}}}

\newcommand{\rhoDotdu}{{\rhoDotElem{\du}}}

\newcommand{\rhouu}{{\rhoElem{\uparrow}}}

\newcommand{\rhodd}{{\rhoElem{\downarrow}}}

\newcommand{\rhoSS}{{\rhoElem{S}}}

\newcommand{\rhoud}{\rhoElem{\ud}}

\newcommand{\rhodu}{\rhoElem{\du}}


\newcommand{\gammaLS}[1]{\gamma_{ls,\, {#1}}}

\newcommand{\sqrtGammaLS}[1]{\Gamma_{ls,\, {#1}}}

\newcommand{\gammaTZero}{\gamma}

\newcommand{\gal}[1]{\gammaTZero_{#1}^{\alpha}}

\newcommand{\gmal}[1]{\gammaTZero_{#1}^{-\alpha}}

\newcommand{\gl}[1]{\gammaTZero_{#1}}

\newcommand{\gul}[1]{\gammaTZero_{#1}^{\uparrow}}

\newcommand{\gdl}[1]{\gammaTZero_{#1}^{\downarrow}}

\newcommand{\gull}{\gul{l}}

\newcommand{\gdll}{\gdl{l}}

\newcommand{\guO}{\gul{1}}

\newcommand{\guT}{\gul{2}}

\newcommand{\gdO}{\gdl{1}}

\newcommand{\gdT}{\gdl{2}}

\newcommand{\gO}{\gammaTZero_{1}}

\newcommand{\gT}{\gammaTZero_{2}}

\newcommand{\Dm}{\Delta\mu}

\newcommand{\HDD}{H_{\mathrm{DD}}}

\newcommand{\HDL}{H_{\mathrm{DL}}}

\newcommand{\tDD}{t_{\mathrm{DD}}}

\newcommand{\tDL}{t_{\mathrm{DL}}}

\newcommand{\sysLbl}{}

\newcommand{\rhoSys}{\rho_{\sysLbl}}

\newcommand{\rhoDotSys}{\dot{\rho}_{\sysLbl}}

\newcommand{\rhoStat}{\bar{\rho}}

\newcommand{\rhoFull}{\rho_{\mathrm{F}}}


\newcommand{\TrB}{{{\mathrm{Tr}}_{\bathLbl}\, }}

\newcommand{\TrS}{{{\mathrm{Tr}}_{\sysLbl}\, }}

\newcommand{\TrF}{{{\mathrm{Tr}}\, }}

\newcommand{\LS}{{L_{\sysLbl}}}

\newcommand{\LBath}{{L_{\bathLbl}}}

\newcommand{\HBath}{{H_{\bathLbl}}}

\newcommand{\LNot}{L_{0}}

\newcommand{\LV}{{L_{V}}}


\newcommand{\decoherenceSymbol}{V}

\newcommand{\rateSymbol}{W}

\newcommand{\effectiveRate}[1]{\rateSymbol_{#1}}

\newcommand{\effectiveRateMax}[1]{W_{#1}^{\mathrm{max}}}

\newcommand{\Xu}{X\uparrow}

\newcommand{\Xd}{X\downarrow}

\newcommand{\XdXu}{\Xd,\Xu}

\newcommand{\XuXd}{\Xu,\Xd}

\newcommand{\XuXu}{\Xu}

\newcommand{\XdXd}{\Xd}

\newcommand{\dXd}{\downarrow,\Xd}

\newcommand{\uXd}{\uparrow,\Xd}

\newcommand{\dXu}{\downarrow,\Xu}

\newcommand{\uXu}{\uparrow,\Xu}

\newcommand{\Xdd}{\Xd,\downarrow}

\newcommand{\Xdu}{\Xd,\uparrow}

\newcommand{\Xud}{\Xu,\downarrow}

\newcommand{\Xuu}{\Xu,\uparrow}

\newcommand{\exLabel}{\mathrm{L}}

\newcommand{\exNot}{X^{0}}

\newcommand{\exM}{X^{-}}

\newcommand{\exMket}{\ket{\exM}}

\newcommand{\exMbra}{\bra{\exM}}

\newcommand{\exMup}{X_{\uparrow}^{-}}

\newcommand{\exMupket}{\ket{\exMup}}
 
\newcommand{\exMupbra}{\bra{X_{\uparrow}^{-}}}

\newcommand{\exMdown}{X_{\downarrow}^{-}}

\newcommand{\exMdownket}{\ket{\exMdown}}
 
\newcommand{\exMdownbra}{\bra{\exMdown}}

\newcommand{\RabiEx}{\Omega_{\exLabel}}

\newcommand{\detuningEx}{\delta_{\exLabel}}

\newcommand{\WeffEx}{\effectiveRate{\exLabel}}

\newcommand{\decohEx}{\decoherenceSymbol_{\mathrm{X}}}

\newcommand{\omegaESR}{\omega_{\mathrm{ESR}}}

\newcommand{\RabiESR}{\Omega_{\mathrm{ESR}}}

\newcommand{\detuningESR}{\delta_{\mathrm{ESR}}}

\newcommand{\WeffESR}{\effectiveRate{\mathrm{ESR}}}

\newcommand{\WmaxESR}{\effectiveRateMax{\mathrm{ESR}}}

\newcommand{\decohESR}{\frac{1}{T_{2}}}

\newcommand{\detuningESReff}{\tilde{\delta}_{\mathrm{ESR}}}

\newcommand{\decohESReff}{\decoherenceSymbol_{\mathrm{ESR}}}

\newcommand{\Wud}{W_{\ud}}

\newcommand{\Wdu}{W_{\du}}

\newcommand{\WSud}{\tilde{W}_{\ud}}

\newcommand{\WSdu}{\tilde{W}_{\du}}

\newcommand{\WXdXu}{\rateSymbol_{\XdXu}}

\newcommand{\WXuXd}{\rateSymbol_{\XuXd}}

\newcommand{\Wem}{\rateSymbol_{\mathrm{em}}}

\newcommand{\Wswitchoff}{\rateSymbol_{\mathrm{r}}}

\newcommand{\bathLbl}{\mathrm{R}}

\newcommand{\HSrot}{H'}

\newcommand{\Hdot}{H_{\mathrm{dot}}}

\newcommand{\HphotonField}{H_{\mathrm{L}}}

\newcommand{\photolum}[1]{\Gamma^{#1}}

\newcommand{\reptime}{\tau_{\mathrm{rep}}}

\newcommand{\rhoinf}{\rho_{\infty}}

\newcommand{\meq}{\mathcal{M}}

\newcommand{\meqL}{\meq_{\mathrm{L}}}

\newcommand{\meqnull}{\meq_{0}}

\title{Optical Detection of Single-Electron Spin Decoherence in a Quantum
Dot }

\author{Oliver Gywat, Hans-Andreas Engel, and Daniel Loss}

\affiliation{Department of Physics and Astronomy, University of Basel, Klingelbergstrasse
82, CH-4056 Basel, Switzerland}

\author{R.J. Epstein, F.M. Mendoza, and D.D. Awschalom}

\affiliation{Center for Spintronics and Quantum Computation, University of California,
Santa Barbara, California 93106, USA}

\begin{abstract}
We propose a method based on optically detected magnetic resonance
(ODMR) to measure the decoherence time $T_{2}$ of a single electron
spin in a semiconductor quantum dot. The electron spin resonance (ESR)
of a single excess electron on a quantum dot is probed by circularly
polarized laser excitation. Due to Pauli blocking, optical excitation
is only possible for one of the electron-spin states. The photoluminescence
is modulated due to the ESR which enables the measurement of electron-spin
decoherence. We study different possible schemes for such an ODMR
setup.
\end{abstract}


\maketitle

\section{Introduction}

Quantum information can be encoded in states of an electron spin $1/2$
in a semiconductor quantum dot.\cite{spintronics} However, information
processing is intrinsically limited by the spin lifetime. For single
spins, one distinguishes between two characteristic decay times $T_{1}$
and $T_{2}$. The relaxation of an excited spin state in a magnetic
field into the thermal equilibrium is associated with the spin relaxation
time $T_{1}$, whereas the spin decoherence time $T_{2}$ is related
to the loss of phase coherence of a single spin that is prepared in
a superposition of its eigenstates. Experimental $T_{2}$ measurements
of single spins in quantum dots are highly desirable because $T_{2}$
is the limiting time scale for coherent spin manipulation. 

Recent optical experiments have demonstrated the coherent control
and the detection of excitonic states of single quantum dots.\cite{gammon1}
Nevertheless, the measurement of the $T_{2}$ time of a single electron
spin in a quantum dot using optical methods has turned out to be an
intricate problem. This is mainly due to the interaction of the electron
and the hole inside an exciton.\cite{current} The electron and hole
spin are decoupled only if the hole spin couples (via spin-orbit interaction)
stronger to the environment than to the electron spin. Recent experiments,
measuring Faraday rotation, have suggested that this is not the case
for excitons in quantum dots.\cite{guptaT2} Alternatively, if electron-hole
pairs are excited inside the barrier material of a quantum dot heterostructure,
the carriers diffuse after their creation to the dots and are captured
inside them within typically tens of picoseconds.\cite{ohnesorge,raymond}
By that time, electron and hole spins have decoupled. In such an experiment,
the Hanle effect would allow the measurement of electron-spin decoherence.
However, this approach\cite{epstein} has not yet given conclusive
results for $T_{2}$. 

What is a promising approach to measure the electron-spin decoherence
time $T_{2}$ by optical methods? For this, initially some coherence
of the electron spin must be produced, preferably in the absence of
holes. This can be done using electron spin resonance (ESR). The coherence
decays and, after some time, the remaining coherence is measured optically.
This implies using optically detected magnetic resonance (ODMR). ODMR
schemes have, e.g., been applied to measure the spin coherence of
single nitrogen-vacancy centers in diamond.\cite{gruber} For quantum
dots, ODMR has recently been applied to electrons and holes in CdSe
dots\cite{Lifshitz} and to excitons in InAs/GaAs dots.\cite{zurauskiene}
While these two experiments have not considered single spin coherence,
the feasibility of the combination of ESR and optical methods in quantum
dot experiments has been demonstrated. 

In this work, we make use of Pauli blocking of exciton creation\cite{pauliblocking}
in an ODMR setup. We show that the linewidth of the photoluminescence
as a function of the ESR field frequency provides a lower bound on
$T_{2}$. Further, if pulsed laser and cw ESR excitation are applied,
electron spin Rabi oscillations can be detected via the photoluminescence. 

We consider quantum dots which confine electrons as \emph{}well as
\emph{}holes (type I dots). We assume a ground state where the dot
is charged with one single electron. This can be achieved, e.g., by
$n$ doping\cite{cortez} or by electrical injection.\cite{abstreiter2}
Such a single-electron state can be optically excited, which leads
to the formation of a negatively charged exciton, consisting of two
electrons and one hole. Recent experiments on InAs dots\cite{trion1,trion2}
and GaAs dots\cite{trion3} have shown that in the charged exciton
ground state, the two electrons form a spin singlet in the lowest
(conduction-band) electron level and the hole occupies the lowest
(valence-band) hole level. Note that single-electron level spacings
can be relatively large, e.g., on the order of 50 meV for InAs dots.\cite{fricke}
Typically, the level spacing of confined hole states is smaller than
the one of electrons.\cite{schmidt} We assume that the lowest heavy
hole (hh) (with total angular momentum projection $J_{z}\!=\!\pm3/2$)
and light hole (lh) ($J_{z}\!=\!\pm1/2$) dot levels are split by
an energy $\delta_{hh-lh}$. Additionally, mixing of hh and lh states
should be negligible.\cite{mixing} These conditions are satisfied
for several types of quantum dots.\cite{trion1,trion2,trion3,efrosCdSe,efros2}
Then, circularly polarized optical excitation that is restricted to
either hh or lh states excites spin-polarized electrons. In this work,
we first assume a hh ground state for holes. We discuss then different
hole configurations. 

\begin{figure}[tb]
 \centerline{\includegraphics[%
  width=8.6cm,
  keepaspectratio]{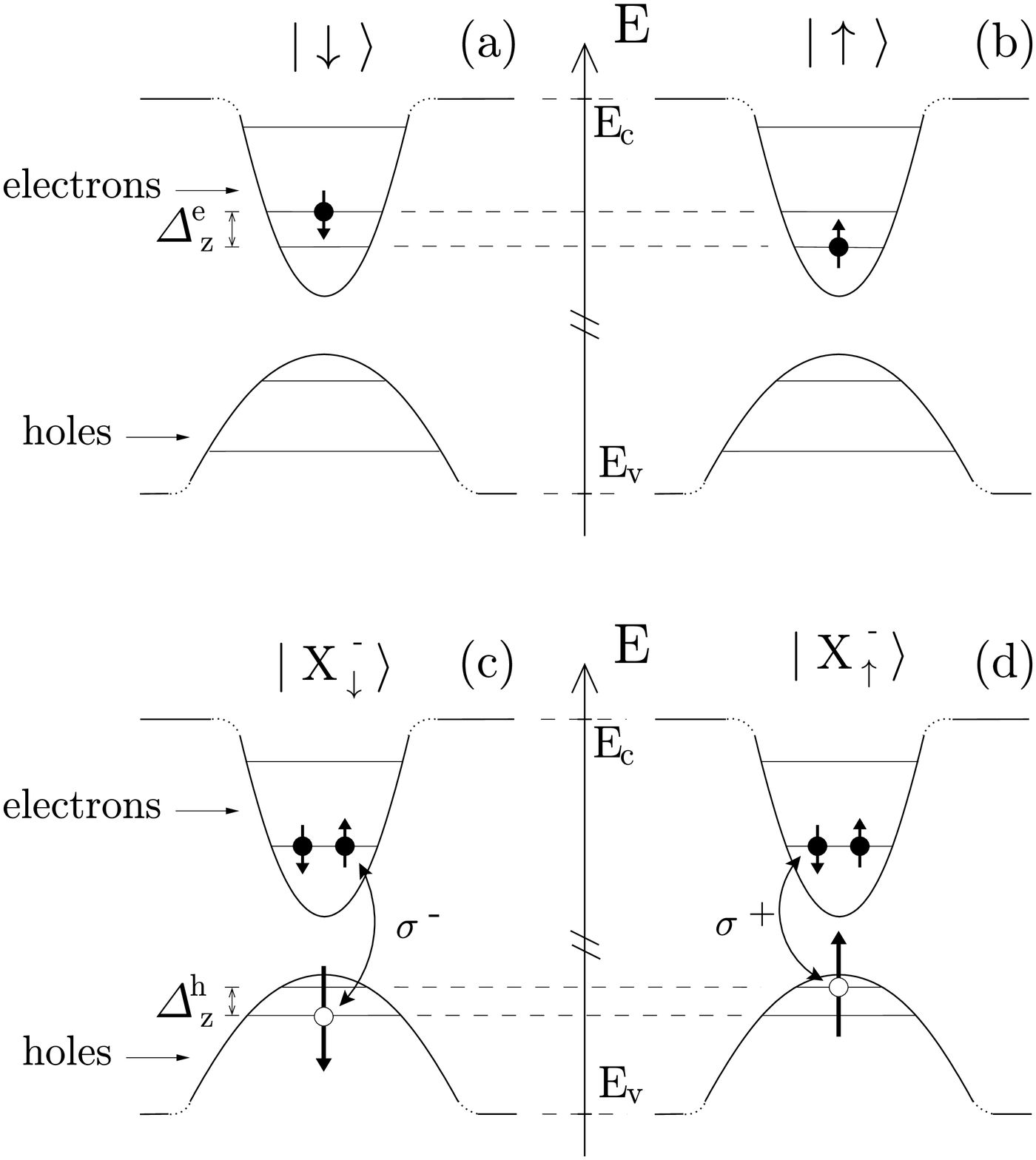} \vspace{1mm} }

\caption{The states of a single quantum dot in a static magnetic field, (a)
$\spdown$, (b) $\spup$, (c) $\exMdownket$, and (d) $\exMupket$.
The Zeeman splittings are $\Delta_{z}^{e}=g_{e}^{z}\mubohr B_{z}$
for the electron and $\Delta_{z}^{h}=g_{hh}^{z}\mubohr B_{z}$ for
the hole. Coherent transitions occur between (a) and (b) due to the
ESR field and between (a) and (c) due to the $\sigma^{-}$-polarized
laser field. The arrows in (c) and (d) indicate which electron-hole
pair couples with the photon field of polarization $\sigma^{\pm}$.
\label{fig:states}}
\end{figure}

The states of a quantum dot can be taken as follows; see also Fig.\ 
\ref{fig:states}. A single electron in the lowest orbital state is
either in the spin ground state $\spup$ or in the excited spin state
$\spdown$. Adding an electron-hole pair, the negatively charged exciton
(in the orbital ground state) is either in the excited spin state
$\exMdownket$ or in the spin ground state $\exMupket$. For these
excitonic states, the subscripts $\downarrow,\uparrow$ refer to the
hh spin and we apply \textbf{}the usual time-inverted \textbf{}notation
for hole spins. For simplicity, we assume $\mathrm{sign}(g_{e}^{z})=\mathrm{sign}\left(g_{hh}^{z}\right)$
for the electron and the hh $g$ factors in $z$ direction. Note that
the very same scheme can also be applied if the sign of $g_{hh}^{z}$
is reversed. Then, one would use a $\sigma^{+}$ laser field and all
results apply after interchanging $\exMdownket$ and $\exMupket$.

\section{Hamiltonian}

We describe the coherent dynamics of a quantum dot, charged with a
single excess electron, in this ODMR setup with the Hamiltonian \begin{equation}
H=\Hdot+H_{\mathrm{ESR}}+\HphotonField+H_{\mathrm{d-L}},\label{eq:H}\end{equation}
 coupling the three states $\spup$, $\spdown$, and $\exMdownket$.
Here, $\Hdot$ comprises the quantum dot potential, the Zeeman energies
due to a constant magnetic field in $z$ direction, and the Coulomb
interaction of electrons and holes. It defines the dot energy $E_{n}$
by $\Hdot\ket{n}=E_{n}\ket{n}$. Here, the electron Zeeman splitting
is $g_{e}^{z}\mubohr B_{z}=E_{\downarrow}-E_{\uparrow}$, where $\mubohr$
is the Bohr magneton.\cite{overhauser}  The ESR term $H_{\mathrm{ESR}}(t)$
couples $\spup$ and $\spdown$ via $\mathbf{B}_{\perp}(t)$, which
rotates with frequency $\omegaESR$ in the $xy$ plane.\cite{linESR,engel}
The ESR Rabi frequency is $\RabiESR=g_{e}^{\perp}\mubohr B_{\perp}$,
with $g$ factor $g_{e}^{\perp}$. Even if the ESR field is also resonant
with the hole Zeeman splitting, it has a negligible effect on the
charged exciton states since they recombine quickly. An oscillating
field $\mubohr\tensor{\mathbf{g}}\mathbf{B}$ can also be produced
with voltage-controlled modulation of the electron $g$ tensor $\tensor{\mathbf{g}}$.\cite{kato}
A $\sigma^{-}$-polarized laser beam is applied in $z$ direction
(typically parallel to $[001]$), with free laser field Hamiltonian
$\HphotonField=\omega_{\exLabel}a_{\exLabel}^{\dagger}a_{\exLabel}$,
where the laser frequency is $\omega_{\exLabel}$, $a_{\exLabel}^{(\dagger)}$are
photon operators, and we set $\hbar=1$. The coupling of $\spdown$
and $\exMdownket$ to the laser field is described by $H_{\mathrm{d}-\mathrm{L}}$
which introduces the complex optical Rabi frequency $\RabiEx$.\cite{optRabi}
Since the dot is only coupled to a single circularly polarized laser
mode via $H_{\mathrm{d-L}}$, the terms that violate energy conservation
vanish due to selection rules. If the laser bandwidth is smaller than
$\delta_{hh-lh}$, the absorption of a $\sigma^{-}$ photon in the
spin ground state $\spup$ is excluded due to Pauli blocking.\cite{trionexc}
We neglect all multi-photon processes via other levels since they
are only relevant to high-intensity laser fields. For this configuration,
the $\sigma^{-}$ photon absorption is switched {}``on'' and {}``off''
by the ESR-induced electron-spin flips. Here, the laser bandwidth
and the temperature can safely exceed the electron Zeeman splitting.
We transform $H$ into the rotating frame with respect to $\omegaESR$
and $\omega_{\mathrm{L}}$. The laser detuning is $\detuningEx=(E_{\Xd}-E_{\downarrow})-\omega_{\mathrm{L}}$
and the ESR detuning $\detuningESR=g_{e}^{z}\mubohr B_{z}-\omegaESR$.

\section{Generalized Master Equation}

We next consider the reduced density matrix for the dot, $\rhoSys=\TrB\rhoFull$,
where $\rhoFull$ is the full density matrix and $\TrB$ is the trace
taken over the environment (or reservoir). In the von Neumann equation
$\dot{\rho}_{\mathrm{F}}=-i\left[H,\,\rhoFull\right]$, we treat the
interaction with the ESR and laser fields exactly with the Hamiltonian
in the rotating frame. We describe the coupling with the environment
(radiation field, nuclear spins, phonons, spin-orbit interaction,
etc.) with phenomenological rates. We write $W_{nm}\equiv W_{n\leftarrow m}$
for (incoherent) transitions from state $\ket{m}$ to $\ket{n}$ and
$\decoherenceSymbol_{nm}$ for the decay of off-diagonal elements
of $\rhoSys$. Note that usually $\decoherenceSymbol_{nm}\geq\frac{1}{2}\sum_{k}\left(W_{kn}+W_{km}\right)$.
The electron-spin relaxation time\cite{kimble} is $T_{1}=\left(\rateSymbol_{\ud}+\rateSymbol_{\du}\right)^{-1}$,
with spin-flip rates $\rateSymbol_{\ud}$ and $\rateSymbol_{\du}$.
In the absence of the ESR and laser excitations, the off-diagonal
matrix elements of the electron spin decay with the (intrinsic) single-spin
decoherence rate $\decoherenceSymbol_{\du}=1/T_{2}$. The linewidth
of the optical $\sigma^{-}$ transition is denoted by $\decohEx=\decoherenceSymbol_{\Xdd}$.
We use the notation $\rho_{n}=\bra{n}\rho\ket{n}$ and $\rho_{nm}=\bra{n}\rho\ket{m}$.
The master equation is given in the rotated basis $\spup$, $\spdown$,
$\exMupket$, $\exMdownket$ as $\rhoDotSys=\mathcal{M}\rhoSys$,
where $\mathcal{M}$ is a superoperator. Explicitly,\begin{eqnarray}
\rhoDotuu & = & \RabiESR{\textrm{Im}}\rhodu\!+\!\Wem\rho_{\XuXu}\!+\!\rateSymbol_{\ud}\rhodd\!-\!\rateSymbol_{\du}\rhouu,\label{eq:fullmastereq}\\
\rhoDotdd & = & -\RabiESR{\textrm{Im}}\rhodu+{\textrm{Im}}(\RabiEx^{*}\rho_{\Xdd})+\Wem\,\rho_{\XdXd}\nonumber \\
 &  & +\rateSymbol_{\du}\,\rhouu-\rateSymbol_{\ud}\,\rhodd,\\
\dot{\rho}_{\XdXd} & = & -{\textrm{Im}}(\RabiEx^{*}\rho_{\Xdd})+\rateSymbol_{\XdXu}\,\rho_{\XuXu}\nonumber \\
 &  & -\left(\Wem+\rateSymbol_{\XuXd}\right)\rho_{\XdXd},\\
\dot{\rho}_{\XuXu} & = & \rateSymbol_{\XuXd}\,\rho_{\XdXd}-\left(\Wem+\rateSymbol_{\XdXu}\right)\rho_{\XuXu},\\
\rhoDotdu & = & \frac{i}{2}\RabiESR\left(\rhodd-\rhouu\right)-\frac{i}{2}\RabiEx^{*}\rho_{\Xdu}\nonumber \\
 &  & -\left(i\detuningESR+T_{2}^{-1}\right)\,\rhodu,\\
\dot{\rho}_{\Xdu} & = & \frac{i}{2}\RabiESR\,\rho_{\Xdd}-\frac{i}{2}\RabiEx\rhodu\nonumber \\
 &  & -[i(\detuningESR+\detuningEx)+\decoherenceSymbol_{\Xdu}]\,\rho_{\Xdu},\\
\dot{\rho}_{\Xdd} & = & \frac{i}{2}\RabiESR\rho_{\Xdu}-\frac{i}{2}\RabiEx(\rhodd-\rho_{\XdXd})\nonumber \\
 &  & -(i\detuningEx+\decohEx)\rho_{\Xdd}.\end{eqnarray}
 The remaining matrix elements of $\rhoSys$ are decoupled and are
not important here. %
\begin{figure}[t]
 \centerline{\includegraphics[%
  width=8.6cm,
  keepaspectratio]{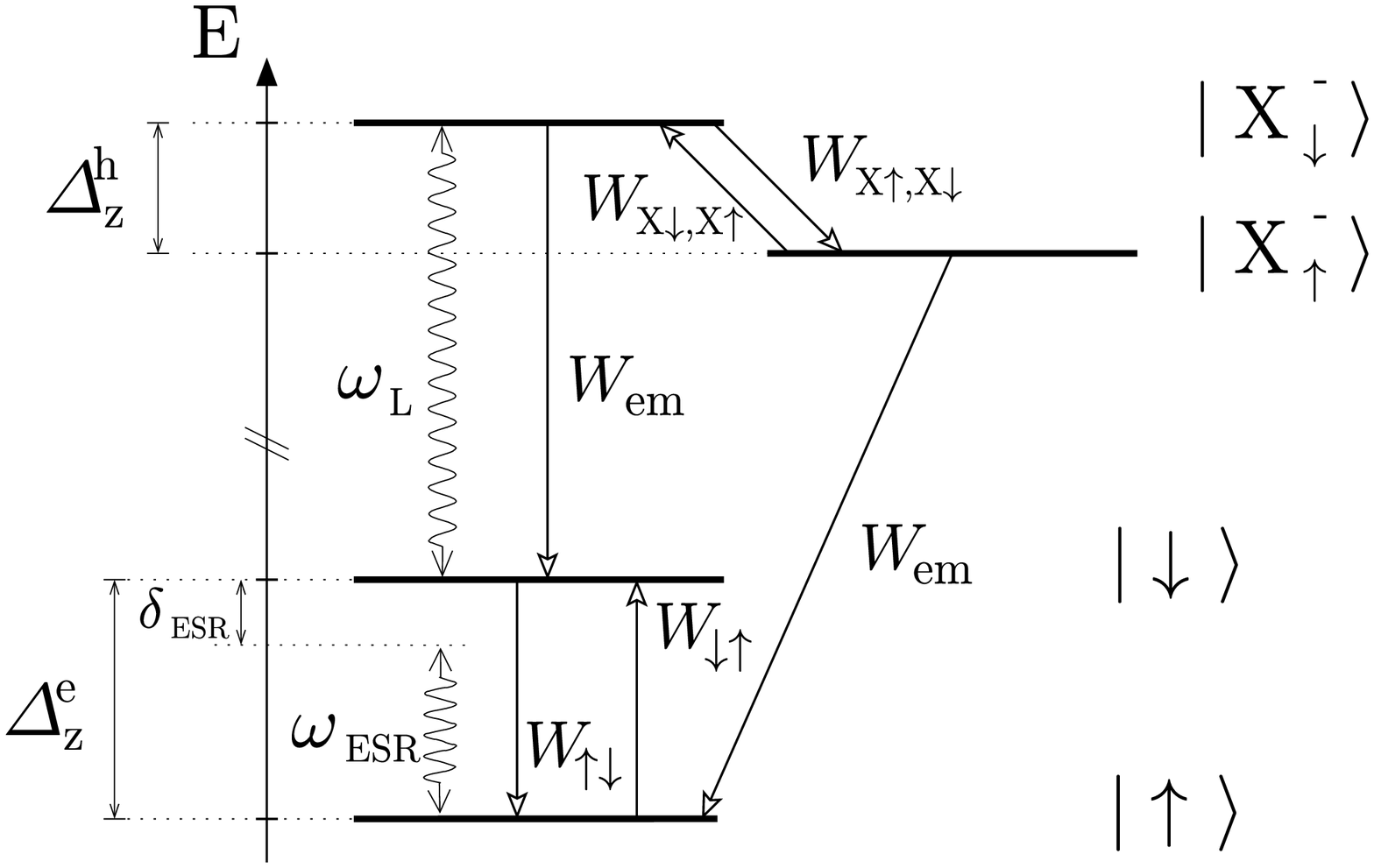} \vspace{1mm} }

\caption{Scheme of the transitions between $\spup$, $\spdown$, $\exMupket$,
and $\exMdownket$. Wavy arrows describe the transitions driven by
the ESR field and the laser field with frequencies $\omegaESR$ and
$\omega_{\exLabel}$, respectively. The corresponding Rabi frequencies
are $\RabiESR$ and $|\RabiEx|$. A detuning $\detuningESR=\Delta_{z}^{e}-\omegaESR$
is shown for the ESR field, with Zeeman splitting $\Delta_{z}^{e}$.
Incoherent transitions are depicted with arrows and occur at rates
$\rateSymbol_{nm}$. We consider $\rateSymbol_{\dXd}=\rateSymbol_{\uXu}=:\rateSymbol_{\mathrm{em}}$.
\label{fig:transitions}}
\end{figure}

\section{ESR Linewidth in Photoluminescence}

We first consider the photoluminescence for a cw ESR and laser field.
For this, we calculate the stationary density matrix $\rhoStat_{\sysLbl}$
with $\dot{\rhoStat}_{\sysLbl}=0$. We introduce the rate \begin{equation}
\WeffEx=\frac{|\RabiEx|^{2}}{2}\:\frac{\decohEx}{\decohEx^{2}+\detuningEx^{2}}\label{eqWeffEx}\end{equation}
 for the optical excitation, with maximum value $\WeffEx^{\mathrm{max}}$
at $\detuningEx=0$. We first solve $\dot{\rhoStat}_{\Xdu}=0$ and
find that the coupling to the laser field produces an additional decoherence
channel to the electron spin. We obtain the renormalized spin decoherence
rate $\decohESReff$ which satisfies\begin{equation}
\decohESReff\leq\decohESR+\frac{|\RabiEx|^{2}}{4\decoherenceSymbol_{\Xdu}}\approx\decohESR+\frac{1}{2}\WeffEx^{\mathrm{max}}.\label{eqndecohESReff}\end{equation}
 Further, the ESR detuning is also renormalized, \begin{equation}
\detuningESReff\geq\detuningESR\left[1-\frac{|\RabiEx|^{2}}{\left(\Wem+\WXuXd\right)^{2}}\right].\end{equation}
 We assume that these renormalizations and $\detuningEx$ are small
compared to the linewidth of the optical transition, i.e., $\WeffEx^{\mathrm{max}},\,|\detuningESReff-\detuningESR|<\decohEx$.
Then, if both transitions are near resonance, $\detuningEx\lesssim\decohEx$
and $|\detuningESReff|\lesssim\decohESReff$, no additional terms
appear in the renormalized master equation. We solve $\dot{\rhoStat}_{\Xdd}=0$
and $\dot{\rhoStat}_{\ud}=0$ and introduce the rate\begin{equation}
\WeffESR=\frac{\RabiESR^{2}}{2}\:\frac{\decohESReff}{{\decohESReff}^{2}+{\detuningESReff}^{2}},\label{eqWeffESR}\end{equation}
which together with $\WeffEx$ eliminates $\RabiEx$, $\decohEx$,
$\detuningEx$, $\RabiESR$, $\decohESReff$, and $\detuningESReff$
from the remaining equations for the diagonal elements of $\rho$.
These now contain the effective spin-flip rates $\WSud=\Wud+\WeffESR$
and $\WSdu=\Wdu+\WeffESR$. %
\begin{figure}[t]
\includegraphics[%
  width=8.6cm,
  keepaspectratio]{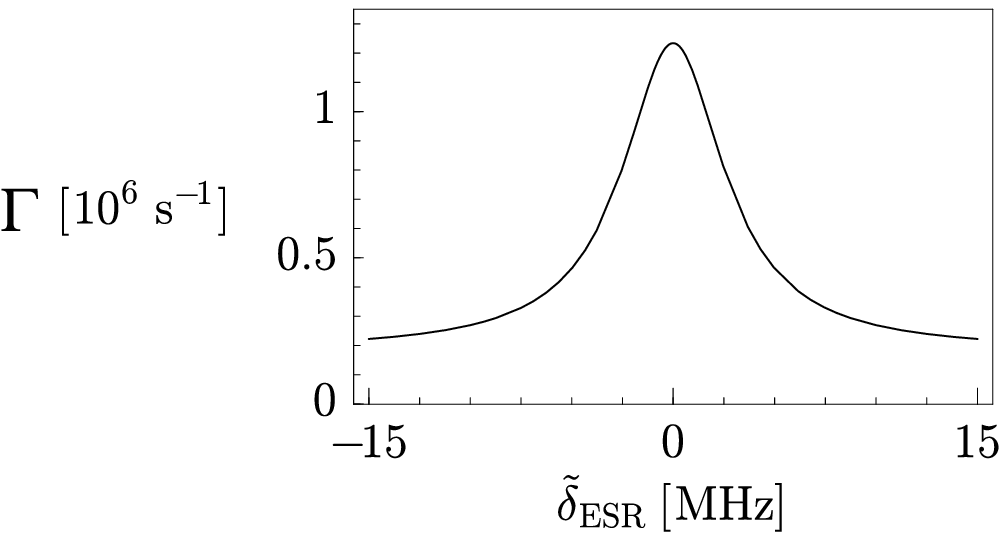}

\caption{\label{Fig: cwPL}The total photoluminescence rate $\Gamma$ is a
Lorentzian as a function of the ESR detuning $\detuningESReff$. Its
linewidth $w$ gives an upper bound for $2/T_{2}$. Here, we use $g_{e}=0.5$,
$B_{\perp}=1\:\mathrm{G}$, $T_{2}=100\:\mathrm{ns}$, $\rateSymbol_{\ud}=\rateSymbol_{\du}=(20\:\mathrm{\mu s})^{-1}$,
$\rateSymbol_{\mathrm{em}}=10^{9}\:\mathrm{s^{-1}}$, $\rateSymbol_{\XuXd}=\rateSymbol_{\XdXu}=\rateSymbol_{\mathrm{em}}/2$,
$\detuningEx=0$, $\decoherenceSymbol_{\Xdu}=\decohEx=(\rateSymbol_{\mathrm{em}}+\rateSymbol_{\XuXd})/2$,
and $\RabiEx=2\RabiESR\sqrt{T_{2}\decohEx}$. With these parameters,
the requirement $\rateSymbol_{\exLabel}\lesssim T_{2}^{-1}\lesssim\decohESReff$
is satisfied.}
\end{figure}
We find the stationary solution\begin{eqnarray}
\rhoStat_{\uparrow} & = & \eta\,\WeffEx\,\Wem\,\WXuXd+\eta\,\WSud\,\Wem\,\WXuXd\nonumber \\
 &  & \;+\eta\,\WSud\left(\WeffEx+\Wem\right)\,\left(\Wem+\WXdXu\right),\qquad\\
\rhoStat_{\downarrow} & = & \eta\,\WSdu\,\left(\WeffEx+\Wem\right)\left(\Wem+\WXdXu\right)\nonumber \\
 &  & \;+\eta\,\WSdu\,\Wem\,\WXuXd,\\
\rhoStat_{\Xd} & = & \eta\,\WeffEx\,\WSdu\,\left(\Wem+\WXdXu\right),\\
\rhoStat_{\Xu} & = & \eta\,\WeffEx\,\WSdu\,\WXuXd,\end{eqnarray}
where the normalization factor $\eta$ is such that $\sum_{n}\rho_{n}=1$.
Note that $\rhoStat_{\uparrow}\geq\rhoStat_{\downarrow}$ is satisfied
for $\Wud\geq\Wdu$. Thus, electron-spin polarization is achieved
due to the hole-spin relaxation channel, analogous to an optical pumping
scheme. Now, photons with $\sigma^{-}$ ($\sigma^{+}$) polarization
are emitted from the dot at the rate $\photolum{-}=\Wem\rhoStat_{\Xd}$
($\photolum{+}=\Wem\rhoStat_{\Xu}$). These rates are proportional
to $\WeffESR/(\gamma+\WeffESR)$ for a given $\gamma$, up to a constant
background which is negligible for $\Wdu<\WeffESR$. In particular,
the total rate $\Gamma=\photolum{-}+\photolum{+}$ as a function of
$\detuningESReff$ is a Lorentzian with linewidth \begin{equation}
w=2\,\decohESReff\sqrt{1+\frac{\WmaxESR}{\gamma}};\end{equation}
 see Fig.\ \ref{Fig: cwPL}. Analyzing the expression for $\gamma$,
we find the relevant parameter regime with the inequality \begin{eqnarray}
w & \leq & 2\,\decohESReff\,\Bigg[1+\frac{2\,\WmaxESR}{\WeffEx}\,\left(1+\frac{\Wem}{\Wswitchoff}+\frac{\WXdXu}{\Wswitchoff}\right)\nonumber \\
 &  & \quad+\frac{3\,\WmaxESR}{\Wswitchoff}+\frac{\WmaxESR}{\Wem}\,\left(1+\frac{3\,\WXdXu}{\Wswitchoff}\right)\Bigg]^{1/2}\!\!\!\!\!\!\!,\label{eqnLinewidth}\end{eqnarray}
which saturates for vanishing $\Wdu$ and $\Wud$. Here, the rate
$\Wswitchoff=\WXuXd+\Wud\left(1+\Wem/\WeffEx\right)$ describes different
relaxation channels, all leading to the ground state $\spup$, and
thus corresponds to {}``switching off'' the laser excitations. If
$\Wswitchoff$ is large, e.g., due to efficient hole-spin relaxation,\cite{Flissikowski}
$w\approx2\,\decohESReff$. From the linewidth $w$ one can extract
a \emph{lower bound for} $T_{2}$: $T_{2}\geq1/\decohESReff\geq2/w$.
Further, this lower bound saturates when the expression in brackets
in Eq.\ (\ref{eqnLinewidth}) becomes close to 1 and $T_{2}^{-1}\approx\decohESReff$
{[}see Eq.\ (\ref{eqndecohESReff}){]}, i.e., the $T_{2}$ time is
given by the linewidth. Comparing with the exact solution, we find
that our analytical approximation gives the value of $\Gamma$ within
$0.2\%$ for the parameters of Fig.\  \ref{Fig: cwPL}. Due to possible
imperfections in this ODMR scheme, e.g., mixing of hh and lh states
or a small contribution of the $\sigma^{+}$ polarization in the laser
light, also the state $\spup$ can be optically excited. We describe
this with the effective rate $\rateSymbol_{\exLabel,\uparrow}$ which
leads to an additional linewidth broadening {[}similar to Eq.\ (\ref{eqnLinewidth}){]}.
This effect is small for $\rateSymbol_{\exLabel,\uparrow}<\WeffESR$.
Detection of the laser stray light can be avoided by only measuring
$\Gamma^{+}$. Otherwise, the laser could be distinguished from $\Gamma^{-}$
by using two-photon absorption. As an alternative, the optical excitation
could be tuned to an excited hole state (hh or lh), possibly with
a reversal of laser polarization. A \emph{pulsed} laser, finally,
would enable the distinction between luminescence and laser light
by time gated detection. %
\begin{figure}[t]
\includegraphics[%
  width=8.6cm,
  keepaspectratio]{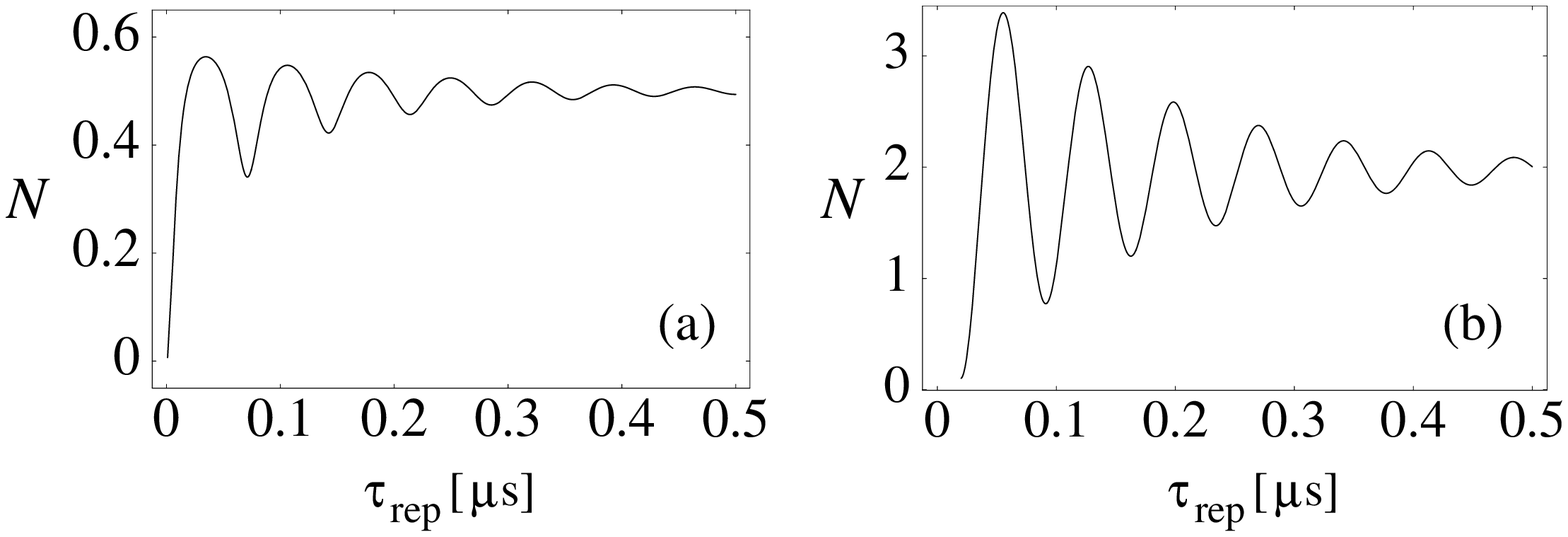}

\caption{\label{Fig: pulsedPL}Average number $N$ of photons emitted per
period $\reptime$ as function of the laser pulse repetition time
for (a) $\pi$ pulses with $\Delta t=5\,\mathrm{ps}$ and $\RabiEx=\pi/\Delta t$,
and (b) pulses with $\Delta t=20\,\mathrm{ns}$ and $\RabiEx=\pi/(500\,\mathrm{ps})$.
We have set $\detuningESR=0$. The other parameters are the same as
in Fig.\ \ref{Fig: cwPL}. The decay of the oscillation is given
by $\decohESReff$ and therefore depends on $T_{2}$.}
\end{figure}

\section{Spin Rabi Oscillations via Photoluminescence}

For a pulsed $\sigma^{-}$ laser, one can also measure $\Gamma$ as
a function of the pulse repetition time $\reptime$ instead of $\detuningESReff$.
We still use cw ESR (or, alternatively, a static transverse magnetic
field, i.e., in the Voigt geometry). We stress that the same restrictions
on the laser bandwidth as in the cw case apply. Due to hole spin flips,
followed by emission of a photon, the dot is preferably in the state
$\spup$ rather than $\spdown$ at the end of a laser pulse. The magnetic
field then acts on the electron spin until the next laser pulse arrives.
Finally, the spin state $\spdown$ is read out optically and, therefore,
the Rabi oscillations (or spin precessions) can be observed in the
photoluminescence as function of $\reptime$; see Fig.\ \ref{Fig: pulsedPL}.
For simplicity, we consider square pulses of length $\Delta t$. We
write in the master equation $\meq(t)=\meqL$ during a laser pulse
and otherwise $\meq(t)=\meqnull$, setting $\RabiEx=0$. We find the
steady-state density matrix $\rhoinf$ of the dot just after the pulse
with $U_{p}\rhoinf=\rhoinf$, where $U_{p}=\exp(\meqL\Delta t)\exp[\meqnull(\reptime-\Delta t)]$
describes the time evolution during $\reptime$. 

The photoluminescence rate is now evaluated by $\Gamma=\Wem\overline{(\rho_{\Xd}+\rho_{\Xu})}$,
where the bar designates time averaging over many periods $\reptime$.
For $\Delta t\geq\pi/\RabiEx,\,\rateSymbol_{\mathrm{em}}^{-1}$, the
spin oscillations become more pronounced; see Fig.\ \ref{Fig: pulsedPL}
(b). This results from an enhanced relaxation to the state $\spup$
during each pulse and thus from a much larger $\rho_{\uparrow}$ than
$\rho_{\downarrow}$ just after the pulse.

\section{Conclusions}

We have proposed an ODMR setup with ESR and polarized optical excitation.
We have shown that this setup allows the optical measurement of the
single-electron spin decoherence time $T_{2}$ in semiconductor quantum
dots. The discussed cw and pulsed optical detection schemes can also
be combined with pulsed instead of cw ESR, allowing spin echo and
similar standard techniques. Such pulses can, e.g., be produced via
the ac Stark effect.\cite{ACstark} Further, as an alternative to
photoluminescence detection, photocurrent can be used to read out
the charged exciton,\cite{abstreiter2} and the same ODMR scheme can
be applied.

\section*{Acknowledgments}

We thank J. C. Egues, A. V. Khaetskii, B. Hecht, and H. Schaefers
for discussions. We acknowledge support from DARPA, ARO, NCCR Nanoscience,
and the Swiss and US NSF.


\begin{thebibliography}{10}
\bibitem{current}Alternatively, one can measure $T_{2}$ via currents through quantum
dots in an ESR field (Refs.\ 24 and 31). However, using an optical
detection scheme there is no need for contacting dots with current
leads (thus reducing decoherence) and one can benefit from the high
sensitivity of photodetectors.
\bibitem{overhauser}In $B_{z}$, we have absorbed the Overhauser field which could possibly
arise from dynamically polarized nuclear spins.
\bibitem{trionexc}Excitation of a charged exciton state containing an electron triplet
would require an additional energy of $\sim40$ meV (Ref.\ 12) and
can be excluded by the low laser bandwidth.
\end{thebibliography}
\end{document}